\documentclass[12pt]{article}
\usepackage{graphicx}
\usepackage{verbatim}

\input epsf
\def\beq{\begin{eqnarray}}
\def\eeq{\end{eqnarray}}

\def\ba{\begin{eqnarray}}
\def\ea{\end{eqnarray}}
\def\beq{\begin{eqnarray}}
\def\eeq{\end{eqnarray}}

\def\p{{\cal P}}

\def\L*{{\cal L}_*}
\def\L{\mathcal{L}}
\def\({\left(}
\def\){\right)}

\def\p{\partial}

\def\lsim{\mathrel{\rlap{\lower3pt\hbox{\hskip0pt$\sim$}}
     \raise1pt\hbox{$<$}}}         
\def\gsim{\mathrel{\rlap{\lower4pt\hbox{\hskip1pt$\sim$}}
     \raise1pt\hbox{$>$}}}         

\def\lsim{\mathrel{\rlap{\lower3pt\hbox{\hskip0pt$\sim$}}
     \raise1pt\hbox{$<$}}}         
\def\gsim{\mathrel{\rlap{\lower4pt\hbox{\hskip1pt$\sim$}}
     \raise1pt\hbox{$>$}}}         

\usepackage{amsmath}
\usepackage{amsfonts}
\usepackage{verbatim}

\begin{document}

\begin{titlepage}
\begin{center}


{\Large\bf  Dynamics of  Unitarization  by Classicalization}

\vspace{0.2cm}

\end{center}

\begin{center}

{\bf Gia Dvali}$^{a,b,c,d}$ and  {\bf David Pirtskhalava}$^{d}$, 


\vspace{.2truecm}

\centerline{\em $^a$Arnold Sommerfeld Center for Theoretical Physics,
Fakult\"at f\"ur Physik} 
\centerline{\em Ludwig-Maximilians-Universit\"at M\"unchen,
Theresienstr.~37, 80333 M\"unchen, Germany}


{\em $^b$Max-Planck-Institut f\"ur Physik,
F\"ohringer Ring 6, 80805 M\"unchen, Germany}


{\em $^c$CERN,
Theory Division,
1211 Geneva 23, Switzerland}


{\em $^d$CCPP,
Department of Physics, New York University\\
4 Washington Place, New York, NY 10003, USA} \\

\end{center}


\centerline{\bf Abstract}

\noindent

 We study dynamics of the classicalization phenomenon suggested in \cite{dggk}, 
 according to which a class of non-renormalizable theories self-unitarizes  at very high-energies  via  creation of classical configurations (classicalons).  We study this  phenomenon in an explicit model  of derivatively-self-coupled scalar that serves as a prototype  for a 
 Nambu-Goldstone-St\"uckelberg field.    We prepare the initial state in form  of a 
 collapsing  wave-packet of a small occupation number but of very high energy, and observe 
 that  the classical configuration indeed develops.  Our results  confirm  the previous estimates, 
 showing that  because of self-sourcing the wave-packet  forms a classicalon configuration with radius that increases with center of mass energy.  Thus,  classicalization takes place before 
 the waves get any chance of probing short-distances. 
   The self-sourcing by energy  is the crucial point, which makes classicalization  phenomenon different from the ordinary dispersion of the wave-packets in 
 other interacting theories.   Thanks to this, unlike solitons or other  non-perturbative objects, the 
 production of classicalons is not only unsuppressed, but in fact dominates the high-energy scattering. 
 In order to make the difference between classicalizing and non-classicalizing theories clear, 
 we use a  language in which the scattering cross section in a generic theory can be universally understood  as a geometric cross section set by a classical radius down to which waves can propagate freely, before being scattered.    We then show, that in non-classicalizing examples this radius  shrinks with increasing energy and becomes microscopic, whereas in classicalizing theories  expands and becomes macroscopic.   
  We study  analogous scattering in a Galileon system and discover that classicalization is less efficient there. We thus observe, that  classicalization is source-sensitive and that Goldstones 
 pass the first test.

\end{titlepage}

\section{Essence of Classicalization}

In the standard (Wilsonian) approach to UV-completion of quantum field theories, it is assumed, 
that non-renormalizable theories with cutoff $M_* \, \equiv L_*^{-1}$ necessarily violate unitarity above the  scale $M_*$, and require its restoration by integrating-in some external  weakly-coupled degrees of freedom.   This approach is based on a common physical intuition, that in a scattering process at higher and higher energies, $\sqrt{s} \, \gg\, M_*$, one is able to probe shorter and shorter  distances, $L \, \sim \, 1/\sqrt{s} \, \ll \, L_*$. 

   It has been suggested recently \cite{dggk}, that  deep-UV picture can be very  different 
in theories that contain extended classical objects ({\it classicalons}) that represent configurations of 
a bosonic field $\phi$ that  is sourced  by  energy (momentum) of the scattering particles.  The characteristic size of the classical configuration, called a classicalization radius $r_*$,  is a model-dependent function of the 
energy of the source $\sqrt{s}$, but the key feature is, that for the trans-cutoff energies 
$\sqrt{s} \, \gg \, M_*$, the classicalization radius exceeds the fundamental length 
$r_*(\sqrt{s}) \, \gg \, L_* $.   An  essential  condition for classicalization is the existence of a classicalizer  field, $\phi$,  which is sourced by the energy-momentum sources  that are  involved in 
a given scattering process.    The role of the classicalon  is to  form an extended classical 
configuration of size  $r_*(\sqrt{s})$, whenever the energy associated with a given process exceeds  the cutoff scale  $M_*$.  In this way, classicalon produces a barrier for probing sub-cutoff distances in a given scattering process\footnote{As pointed out in \cite{dggk},  $r_*$-radius plays the role somewhat analogous to Schwarzschild radius for  non-gravitational theories, whereas 
classicalons share some obvious analogy  with black holes, in the sense that they both represent  
strong sources of the corresponding classicalon fields.}.

    The consequences of such a picture  would be pretty profound.  Because of classicalization, it becomes impossible to localize energy at distances shorter than the corresponding $r_*$-radius. 
  As a result, the  $2\rightarrow 2$ particle scatterings  at large center of mass energy $\sqrt{s}\gg M_*$ happen  only through very low momentum-transfer $\sim r_*^{-1}$, and give negligible contribution to the scattering process.  Instead, the cross section is dominated by creation of classical configurations  that decay into many particle states. 
  
   From the first glance, such a scenario may look puzzling, since creation of classical objects, such as topological solitons,   in two-particle collisions  must be {\it exponentially-suppressed}.
   So why are the classicalons created? 
   The answer is, because  classicalons  are  {\it not}  solitons. They are configurations of the 
   $\phi$-field  sourced by the energy, which is a  {\it Noether}-type  rather than a topological-type charge.  Therefore, appearance of a classicalon configuration  is inevitable whenever 
localization of high-enough energy takes place.  Correspondingly, in processes that exhibit classicalization,  it is simply impossible  to form two-particle quantum states with energy $\sqrt{s}\, \gg \ M_*$, localized 
  at distances  $\ll r_*(\sqrt{s})$!  Any such localized state is automatically a classical object and thus,  represents a many-particle state.  In other words,  when preparing  the scattering experiment
  by bringing two highly  energetic particles within the impact parameter  
from an initially-infinite separation,  we are gradually building a classical configuration via sourcing the classicalizer field
  by the center of mass energy of the  colliding particles. By the time particles come  within  the $r_*(\sqrt{s})$-radius corresponding to their center of mass energy,  the system is already classical  and no 
  longer represents a two-particle quantum state.   So effectively the only chance for
  $2\rightarrow 2$ scattering to take place is,  if particles never come closer than $r_*$-distance. 
  Thus,  in $2\rightarrow 2$ particle scattering, 
  the momentum transfer is limited by  $r_*(\sqrt{s})^{-1}$. 
  
   In other words, classicalon configurations are a necessary consequence  of localized 
   energy, because the latter sources the classicalizer field, and system has no other choice, but 
   to produce one. 
      This goes in a sharp difference with solitons or other non-perturbative objects that represent very special coherent states among the 
   exponentially-large number of possible states  compatible with the same energy.  This is why in 
   two (or few) particle collisions production of the latter objects is exponentially-suppressed even at very high energies \cite{solitons}.

    
     Obviously, there are many open questions, most pressing of which is, does classicalization really work?  
   
   In this note, following \cite{dggk}, we shall take a small step towards understanding the dynamics of classicalization.   Our idea is to prepare an initial wave-packet  with very high-energy 
   $\sqrt{s} \gg M_*$  but a relatively small occupation number, and investigate how the classical configuration of radius  $r_*(\sqrt{s}) \, \gg \, L_*$  develops in the scattering process. 
   
    In order to better understand physics behind the classicalization phenomenon, we shall first confront  classicalizing theories with the ones that do not exhibit this phenomenon. 
 For a better comparison we shall develop an universal language,  in which the scattering cross section  in a generic theory can be understood as a {\it geometric } cross-section defined by the radius  $r_*$. 
 The physical meaning  of the $r_*$-radius  is of a  shortest distance down to which in a given scattering process  waves can propagate freely, 
 without experiencing a significant  interaction.   In this language,  the scattering cross section is universally given by,  
 \begin{equation}
  \sigma \,  \sim \, r_*(\sqrt{s})^2 \, . 
  \label{unisection}
  \end{equation}
  This  relation holds  true both in weakly-coupled examples that do not classicalize, as well as in classicalizing theories.   For example,  for Thomson scattering, the role of $r_*$ is played 
  by the classical radius of electron.   
  
   However, what distinguishes the two cases is  the dependence of $r_*$-radius  on energy. 
In non-classicalizing theories  $r_*(\sqrt{s})$  {\it diminishes}  with growing  $\sqrt{s}$, whereas in classicalizing  theories it {\it grows}  and eventually becomes macroscopic. 
 As a result, in the former theories particles can probe shorter and shorter distances 
 ($\sim \, 1/\sqrt{s})$, before being scattered, whereas in classiclizing theories  scattering takes place already  at $r_*$.

    The analogy between  $r_*$ and the gravitational Schwarzschild radius is clear.  We can define a classical 
 Schwarzschild radius for an arbitrary quantum particle, but a particle  for which the Schwarzschild radius is deep within its Compton wavelength cannot be regarded  as  a classical
 state.   
 Similarly, the classical $r_*$-radius can be defined for an arbitrary wave-packet, but only the ones with  $r_* \, \gg \, L_*$  classicalize.

  We shall examine the two systems.  The first is a theory of a derivatively   self-interacting scalar that serves as a prototype of the Nambu-Goldstone boson or equivalently of longitudinal 
  (St\"uckelberg) component of a massive vector field.  We study $S$-wave scattering and observe that the configuration indeed classicalizes due to self-sourcing of the wave-packet. 
   Both, for monochromatic waves as well as for sharply-localized wave-packets, we  
 find that the  $r_*$-radius  is given by, 
 \begin{equation}
   r_*(\sqrt{s}) \, = \, L_* (L_*\sqrt{s})^{{1\over 3}} \, , 
   \label{centralstar}
   \end{equation}
which is in agreement with the estimates of \cite{dggk}.    Corresponding $2\rightarrow 2$ 
scattering amplitude then  is estimated as, 
 \begin{equation}
 A_{2\rightarrow 2} \sim \, (L_*\sqrt{s})^{-{4 \over 3}} \, ,  
 \label{amplitudefirst}
 \end{equation}    
whereas the total cross-section (dominated by the production of classical configuration) is given 
by, 
 \begin{equation}
  \sigma \,  \sim \,  L_*^2 (L_*\sqrt{s})^{{2\over 3}} \, .
  \label{unisectiongold}
  \end{equation}
  All the above indicates that Goldstone-type derivatively-coupled  scalars are interesting candidates for classicalization.   Of course,  as explained in \cite{dggk}, for classicalization to work, it is  essential that there exist no light  (with mass $\ll \,  M_*$) weakly-coupled radial degree of freedom, otherwise system never classicalizes, and instead, takes the conventional root of a weakly-coupled perturbative unitarization.  
  
    We  next explore the situation for another scalar theory, which describes  a certain decoupling limit  of a generally-covariant theory of a massive resonance graviton (DGP gravity\cite{dgp}). 
    In this limit  the only remaining interacting component of the graviton  is  a helicity-zero St\"uckelberg component of the massive spin-2 state.    We discover that unlike the Goldstone case, in the latter  theory in the  
  $S$-wave scattering classicalization does not happen in the leading order in
  self-coupling. The reason can be traced to the Galilean  symmetry \cite{lpr}  that  requires an excessive number of derivatives in the self-coupling of the scalar.    
  
   We confront derivatively-(self)coupled bosons with other types of self-interacting 
 scalars and show the crucial  role of energy self-sourcing for classicalization.   
  Since energy is conserved,  the scalars that couple to energy density  necessarily get
  a very strong source, whenever energy is localized, and classicalize.  This  does not 
  hold if the source is not of the energy-type.   Classicalization  thus  is different from 
  usual dispersion of the wave-packet in self-interacting theories.

   We thus observe, that classicalization  is sensitive to the nature of self-sourcing, and that 
  Nambu-Goldstone-type scalars  pass the first test. 
   
\section{Scattering with and without Classicalization}

    We wish to give a preliminary discussion of the classicalization phenomenon in a scattering 
 process.    In order to understand the peculiarity of a classicalizing scattering,  we wish to confront it with  scattering in  theories where classicalization does not happen. 
 
  In order to achieve a maximally clear comparison, it is useful to reduce the derivation of the cross section   in both classes of theories to a common language, in which the cross section can be understood as {\it geometric}  cross-section set by $r_*$-radius,  a shortest distance  
till which waves can propagate freely, before being scattered by interaction.  
With this definition, the concept of $r_*$-radius extends to weakly-coupled 
non-classicalizing theories. This extension allows us an universal treatment.

 We shall then see, that in non-classicalizing weakly-coupled theories this distance 
 diminishes with energy, whereas in the classicalizing theories 
 grows.   This is the fundamental difference between the theories 
 that classicalize from the  ones that do not exhibit such a phenomenon. 
 
\subsection{Understanding cross-section as geometric cross-section}

  As the first  example of non-classicalizing theory,  we shall  consider a theory of massless scalar field with a renormalizable  self-interaction, 
   \begin{equation}
 \mathcal{L}= {1\over 2}  \(\partial_{\mu}\phi\)^2 \, - \, {\lambda \over 4}  \phi^4 \, .
 \label{nambu}
  \end{equation}
  At the weak coupling, $\lambda \ll 1$, the $2\rightarrow 2$ perturbative scattering amplitude 
  in the above theory goes as 
  \begin{equation}
   A_{2\rightarrow 2} \sim \lambda \, ,
   \label{phi4amplitude}
   \end{equation}
   and the cross section goes as
  \begin{equation}
  \sigma_{2\rightarrow 2} \, \sim \, {\lambda^2 \over s}  \, .
  \label{crossweak}
  \end{equation}
   In this theory  a tree-level  scattering can probe arbitrarily-short distances, and classicalization does not happen.   In order to understand this fact, let us try to understand 
   the behavior of the scattering amplitude and of the cross-section  in the language of 
   solving the  equation of motion, 
  \begin{equation}
  \Box\phi \,   =  - \lambda \, \phi^3 \, , 
     \label{equphi4}
  \end{equation}
for scattering wave-packets.   We shall assume that for $r \, = \, \infty$ and $t  \, = \, - \infty$, 
$\phi$ is well-approximated by a spherical wave of frequency $\omega$, and 
the amplitude $A \sim 1$ (small occupation number),  
\begin{equation}
\phi_0 \, = \, {\psi(\omega(r+t))  \over r},
\label{wavepacketfree}
\end{equation}
which solves the free-field equation of motion,  
  \begin{equation}
  \Box\phi_0 \,   =  \, 0  \, . 
     \label{equfree}
  \end{equation}
We shall now solve the equation (\ref{equphi4})  iteratively, by expanding the $\phi$-field in series of sub-leading corrections,  
\begin{equation}
\phi \, = \, \phi_0 \, + \, \phi_1 \, + \, ... \, , 
\label{zeroone}
\end{equation}
and trying to understand at what distances  the correction $\phi_1$ to a free-wave becomes 
significant.   For this we have to solve the equation, 
  \begin{equation}
  \Box\phi_1 \,   =  - \lambda \, \phi_0^3 \, =  -\lambda \, {\psi(\omega(r+t))^3  \over r^3} \, .
     \label{equphi4-1}
  \end{equation}
  The solution of this equation, for  $\omega  \, \gg \, r^{-1}$,  behaves as, 
  \begin{equation}
  \phi_1 \, \simeq \, -\frac{\lambda}{2}\,  {\int_0^{r+t} \psi(\omega y)^3 dy \over r^{2}} \, .
  \label{aprsol1}
  \end{equation}
 Notice, that  since  $\psi(wy)$ is a periodic function with amplitude $\sim 1$ and frequency 
 $\omega$,   we have (on average),   
 \begin{equation}
\int_0^{r+t} \psi(\omega y)^3 dy  \, \sim \, {1 \over \omega} \psi(w(r+t))  \sim {1\over \omega} \, .  
  \label{relationlambda}
  \end{equation}
  For example, taking  $\psi \, = \, \cos(\omega(t+r))$ we get, 
  \begin{equation}
  \phi_1 \, = \,  -\frac{\lambda}{6}\,  {\sin\(\omega(t+r)\) ( 2 \, + \, \cos^2 (\omega(t+r))) \over \omega r^{2}} \, .
  \label{equation}
  \end{equation}
 Thus we find,  
  \begin{equation}
  \phi_1 \, \sim \, {\lambda\, \over \omega r} \phi_0 \, .
  \label{omega1}
  \end{equation}
This relation plays the central role in deriving the scattering cross-section in $\phi^4$-theory, since it shows, that  the wave  propagates freely  till the distance, 
\begin{equation}
  r_* \, \sim \,  {\lambda \over \omega} \, .
  \label{rzero}
  \end{equation}
  What is the significance  of the above relation? 
  It tells us, that the high-frequency modes go deeper in UV before being disturbed by the interaction.    
 Thus, the scattering becomes less and less significant  for larger and larger energies. 
 However at  $r \sim r_*$, the correction to the free-wave 
 becomes order one, and the scattering takes place. The scattering cross section thus can be understood as a {\it geometric}  cross section, 
 \begin{equation}
 \sigma \, \sim \, r_*^2 \, \sim \,  {\lambda^2 \over \omega^2} \, , 
 \label{rzerocross}
 \end{equation}
  which reproduces  (\ref{crossweak}).   The above cross section diminishes at high energies 
  because the energetic waves probe shorter and shorter distances.  
 Let us confront the above situation with a theory in which the field is self-sourced by the energy.
 As we shall see, the outcome is very different there.  
 
 \subsection{Classicalization in Goldstone Scattering} 
 
  The example now we wish to consider is of  a scalar  field $\phi$ with the derivative self-interactions,      
   \begin{equation}
 \mathcal{L}= {1\over 2}  \(\partial_{\mu}\phi\)^2 \, + \, {L_*^4\over 4}  \((\partial_{\mu}\phi)^2\)^2 \, .
 \label{nambu}
  \end{equation}
  This theory is symmetric  under the shift by an arbitrary constant $c$,   
  \begin{equation}
  \phi \, \rightarrow \, \phi \, + \,  c \, , 
\label{goldstone}
\end{equation}  
and therefore represents  a  simple prototype describing a self-interacting Nambu-Goldstone field $\phi$, or 
a longitudinal (St\"uckelberg) component 
of a massive vector field.  
 As shown in  \cite{dhk}, the above theory can be regarded as the decoupling limit of  the 
 self-interacting theory of a massive Proca vector field $W_{\mu} \, \equiv \, \tilde{W}_{\mu} \, - \, m_W^{-1} \partial_{\mu} 
 \phi$, with the following Lagrangian
  \begin{equation}
  {\cal L} \, =\,  -{1\over 4} F_{\mu\nu}F^{\mu\nu} \, + \, {1\over 2} m_W^2\, W_{\mu}W^{\mu}  \,   + \, 
  {g^4 \over 4} \,  \, (W_{\mu}W^{\mu})^2 \, .
  \label{procaint}
  \end{equation}
The Lagrangian (\ref{nambu})  is obtained from (\ref{procaint}) in the limit  $g \rightarrow 0$ with $L_*\equiv  g/M_W =$fixed.  In this limit, the transverse gauge field decouples 
  and we are left with a self-interacting  Nambu-Goldstone-St\"uckelberg mode, $\phi$. 
 This theory is known to classicalize  for the localized external probe sources.  We wish to 
 study classicalization  in a simplest  scattering process. 
 
   Perturbatively,  at  the frequency $\omega$, the amplitude  of the $2\rightarrow 2$ scattering process in the above theory goes as  $A_{2\rightarrow 2} \, \sim \omega^4L_*^4$, and naive perturbative cross section 
   is dominated by the high-momentum transfer  processes, growing as 
  \begin{equation}
  \sigma_{2\rightarrow 2} \, \sim \, {L_*^8 \omega^6} \, ,   
  \label{omegalarge}
  \end{equation}
 thus, violating unitarity for $\omega \, \gg \, M_*$.   We now wish to check if this conclusion 
 holds for non-perturbative analysis. 
  
 For this we shall study the scattering of waves by directly analyzing  the equation of motion
 following from (\ref{nambu}), which  reads as follows  
  \begin{equation}
  \partial^{\mu} (\partial_{\mu}\phi \(1+ L^4_*(\partial_{\nu}\phi)^2)\) = 0 \, . 
  \label{goldequation}
  \end{equation}
As in the case of analyzing scattering in $\phi^4$-theory,  we shall assume that for $r \, = \, \infty$ and $t  \, = \, - \infty$, 
$\phi$ is well-approximated by a spherical wave of frequency $\omega$, and 
the amplitude $A \sim 1$ given by (\ref{wavepacketfree}), and solve the equation 
iteratively by performing the  expansion (\ref{zeroone}).  The equation for 
the leading correction to the free wave now becomes, 
 \begin{equation}
  \Box\phi_1  =  - L^4_*
     \partial^{\mu} (\partial_{\mu}\phi_0 (\partial_{\nu}\phi_0)^2) \, .
     \label{phione1}
  \end{equation}
Taking into the account properties of $\psi(\omega(t+r))$-wave,  for 
$\omega \, \gg \, r$, the leading contribution to the right hand side is,  
 \begin{equation}
  \Box\phi_1  =  -  {L^4_*  \over r^5} \, ( 2 \psi^2\psi '' \, + \, 8 \psi\psi '^2) \, ,
     \label{phionea}
  \end{equation}
where prime denotes the derivative with respect to the argument. 
Again, for $\omega \, \gg \, r^{-1}$ the solution of this equation can be approximated by, 
  \begin{equation}
  \phi_1 \, \simeq \, - f(\omega(r+t)) \,  \frac{L_*^4 }{6 r^{4}} \, ,  
  \label{aprphi1}
  \end{equation}
  where,  
 \begin{equation}
 f(\omega(r+t)) \, \equiv  \, \int_0^{r+t} ( 2 \psi^2\psi '' \, + \, 8 \psi\psi '^2) dy \, .
 \label{fdefinition}
 \end{equation}  
 Notice, that  since  $\psi(wy)$ is a periodic function of amplitude $\sim 1$ and frequency 
 $\omega$,   we have,   
 \begin{equation}
f \, \sim \,  \,  \omega \psi \, \sim \, \omega \, . 
  \label{omegaplus}
  \end{equation}
  This relation can be readily checked on an explicit form of $\psi$. For example, for 
  $\psi \, = \, \cos (\omega (t+r))$ we have, $f \, = \,\omega \( 3 \sin (\omega (t+r))-5\sin (3\omega (t+r))\)/6$.
  
  Thus,  we obtain the following relation between the initial free-wave and the leading perturbation due to scattering, 
  \begin{equation}
  \phi_1 \, \sim \, \left ({r_* \over  r} \right )^3 \phi_0 \, ,
  \label{starrelation}
  \end{equation}
where, 
\begin{equation}
r_* \equiv  L_* (\omega L_*)^{1 \over 3} \, .
 \label{radstar} 
 \end{equation}
 This equation tells us, that unlike the $\phi^4$-case, the scattering of Goldstones starts 
already  at a distance $r_*$, which grows at large  $\omega$.   
  For $\omega \, \gg \, L_*$,  the $2\rightarrow 2$ particle scattering is thus dominated by a very low momentum transfer $\sim \, r_*^{-1}$ ,  and must go  as 
 \begin{equation}
 A_{2\rightarrow 2} \sim \, (L_*/r_*)^4 \, \, . 
 \label{trueamplitude}
 \end{equation}    
Thus,  scattering softens at high energy.  This  is a consequence of classicalization.  
 Because of energy self-sourcing,  the initial wave adiabatically re-scatters and  by the time it reaches $r_*$, the correction to a free-wave becomes important.  
High energy scattering becomes governed by a long-distance physics.

\section{Classicalization of Localized Goldstone  Wave-Packets}

 We have seen, that  scattering of monochromatic  waves leads to the formation of 
 $r_*$-radius, which indicates, that with growing energy, waves  tend to scatter before reaching short distances.  Thus two-to-two particle scatterings  must be dominated by very low 
 momentum-transfer.  The high  momentum-transfer scattering on the other hand should be  accompanied  by formation of  a configuration of  {\it classical} radius $r_*$. The system classicalizes at high 
 $\sqrt{s}$.   We wish to show that this property persists for the sharply localized wave-packets.

 Thus, we shall now  consider classicalization of sharply-localized wave-packets.  
  We shall limit our analysis to studying classicalization  in $S$-wave scattering. We wish 
to see if  and how the energetic wave-packets "refuse" to get localized and develop  
the $r_*$-radius.    For this, we shall prepare the initial state of $\phi$ in the form of a collapsing spherical wave-packet  of radius $r$, amplitude $A$,  and a characteristic thickness $a$,  
\begin{equation}
\phi_0 = {\psi((r+t)/a) \over r},
\label{zerophi}
\end{equation}
where $\psi(x/a) $ represents a sharply localized function of  width $a$ around $x=0$.
 For small $A$ and small $a$ the initial wave-packet can be considered as a quantum state with low occupation number,  which however may have an arbitrarily  high energy, $\sqrt{s} \sim A^2/a$.

 The spherical wave-packet  starts at $t = - \infty$ at $r=\infty$,  and collapses  towards the origin. We shall study its spread-out in the course of this evolution.  
  For this we shall solve the equation (\ref{goldequation}) iteratively, by building up perturbations  in powers of 
  $L_*^4$ around the solution $\phi_0$ of the linearized equation,  $$\phi \, = \, \phi_0 \, + \phi_1\, + ...~.$$
  As we shall see, as long as $a < r$,    $\phi_1\, \sim \, {A^3 L_*^4\over  ar^4}$ and  
   the role of a small expansion parameter is therefore played by the following  quantity 
   \begin{equation}
   \epsilon \, \equiv \, A^2L_*^4/(ar^3)  .
   \label{parameter}
\end{equation}   
      
 Thus, the first order-in-$\epsilon$  correction to the free, collapsing wave-packet can be found as 
 the solution of the following equation,  
 \begin{equation}
  \Box\phi_1  =  - L^4_*
     \partial^{\mu} (\partial_{\mu}\phi_0 (\partial_{\nu}\phi_0)^2) \, ,
     \label{phione1}
  \end{equation}
  where the $\phi_0$-dependent part acts as a source for $\phi_1$.   
 
  We shall solve this equation in two different ways.  
  
  First, in order to develop an intuition, we shall replace the source by a time-dependent point-like source of the strength that at each $t$ equals  to a spatially-integrated value of the exact source. 
  
  Later, we shall solve the equation exactly without the former approximation and observe that to the leading order the two procedures give the solutions with identical asymptotic behavior, up to a  multiplicative factor $-4/3$, which is consistent with the averaging procedure.  

\subsection{Classicalization from the averaged source} 
  
   We shall discuss the method of the average source first. 
  Since for any given $t$, $\phi_0$ is localized within the radius $r \, \sim \, |t|$, we can integrate the source over a sphere of radius $ \gg |t|$ and  at distances  $r \gg  |t|$  use  an effective equation with a delta-function type source ,
 \begin{equation}
  \Box\phi_1  =   \delta(\vec{r}) Q(t) \,. 
\label{deltasourcephi}
\end{equation}
Here
\begin{equation}
 Q(t) \, = -L^4_*\, \int_{R \gg |t|}  d^3x \, \partial^{\mu} (\partial_{\mu}\phi_0 (\partial_{\nu}\phi_0)^2) \, , 
 \label{charge}
  \end{equation}
  where volume integration is performed within a sphere of radius $R \gg |t|$. 
 Since $\phi_0$ is a localized function, the integral of the spatial divergence
\begin{equation}
  \int_{R\gg |t|}  d^3x \, \partial^{j} \(\partial_{j}\phi_0 (\partial_{\nu}\phi_0)^2\) \, = \, 0\, ,  
  \end{equation}
  is the surface flux, 
which  vanishes  through any sphere outside the wave-packet.  Thus,  we are left with the following 
effective source,  
 \begin{equation} 
Q(t)\, = -L^4_* \, \int d^3x  \partial_t \(\partial_t\phi_0(\partial_{\nu}\phi_0)^2\) \,  .  
  \end{equation}
Taking into the account the form of the zeroth order solution (\ref{zerophi}), 
  we get, 
 \begin{equation} 
Q(t)\, = 4 \pi L^4_*  \, \int dr  \( \,{(\psi'\psi^2)' \over r^3}- {(2\psi \psi^{'2})' \over r^2} \, \), 
\end{equation}
where prime represents a derivative with respect to $r+t$ (which is the same as $\partial_r$ or $\partial_t$).
After a few partial integrations, we obtain, 
 \begin{equation} 
Q(t)\, =  \, 16\pi L^4_* \int dr  \( \, {\psi^3 \over r^5} -{(\psi \psi^{'2}) \over r^3}\, \)\, .
\label{source''}
\end{equation}
Let us now assume a particular form of the zeroth-order solution, 
\begin{equation}
\phi_0 \, = \, {\psi(r+t)  \over r} \, = \,  A \, {e^{-{(r+t)^2 \over  a^2}} \over r} ,
\label{wavepacket}
\end{equation}
which represents a wave-packet  of amplitude  $A$ and energy $\sim A^2/a$. 
Using this expression, we can evaluate the source in (\ref{source''})
\beq
Q(t)\, =  -\, 16\pi L^4_* A^3 \int dr \(\frac{2}{3}\frac{1}{a^2r^3}+\frac{1}{9 r^3}\p^2_r-\frac{1}{r^5}\)e^{-\frac{3(r+t)^2}{a^2}}.
\eeq
We will now take the limit of an infinitely-strongly localized wave-packet $a\to 0$, so that the last expression becomes
\beq
Q(t)\, =  -\, 16\pi L^4_* A^3 \sqrt{\frac{\pi}{3}}\int dr \(\frac{2}{3}\frac{1}{a r^3}+\frac{a}{9 r^3}\p^2_r-\frac{a}{r^5}\)\delta(r+t).
\eeq
The first term on the right hand side of the latter equation, being enhanced by inverse of the wave-packet width, represents the leading contribution to the effective source for $|t|\gg a$, 
\beq
Q(t\ll -a)= \frac{32 \pi}{3} \sqrt{\frac{\pi}{3}} L^4_* A^3\frac{1}{a t^3}.
\eeq
 

The solution of (\ref{deltasourcephi})  with the above  source gives the spread of the wave-packet for $r\gg|t|$
  \begin{equation}
\phi_1 \, = \,  {Q(t -r) \over 4 \pi r}  =  \frac{8 }{3} \sqrt{\frac{\pi}{3}} L^4_* A^3\frac{1}{a (t-r)^3 r} .
 \label{phione}
 \end{equation}
 This solution indicates that an energetic wave-packet of $\phi$  (with $\sqrt{s} \sim A^2/a \, \gg M_*$)  classicalizes,  since  the  value of the gradient reaches $M_*^2$ at distances $r_* \gg L_*$. 
  We shall come back to a more precise estimate of $r_*$ shortly.  
 
   As a next step, we shall obtain the exact solution of (\ref{phione1}), without averaging the source.

  \subsection{Exact solution}

Below we will solve the equation (\ref{phione1}) explicitly, by taking the same initial-state wave-packet $\phi_0$ as before, and constructing the perturbative solution. The details of derivation of the perturbative solution are given in the appendix. Here we shall reproduce only the 
essential steps. 

Let us introduce the exact source $j(r,t)$, defined by 
\beq
 j(r,t)\equiv \Box \phi_1 = -L^4_*~ \p^\mu (\p_\mu \phi_0 (\p_\nu \phi_0)^2)\,.
\label{eq'}
\eeq
Taking (\ref{zerophi}) into the account, the source can be rewritten as,  
\beq
 j(r,t)=-L^4_* \left( 2 \frac{\psi^2\psi ''}{r^5}+8 \frac{\psi\psi '^2}{r^5}-12 \frac{\psi^2\psi '}{r^6}+4 \frac{\psi^3}{r^7} \right).\
\label{source'}
\eeq
Evaluating the latter expression for the zeroth order wave-packet $\psi=A e^{-\frac{(r+t)^2}{a^2}} \,$, and later taking the $a\to 0$ limit as above, $j(r,t)$ can be recast in the following form
\beq
j(r,t)=- L^4_* A^3 \sqrt{\frac{\pi}{3}} \(\frac{8}{3 a r^5} -\frac{4a}{r^6}\p_r +\frac{10a}{9r^5} \p^2_r  +\frac{4a}{r^7} \) \delta(r+t).
\eeq
The equation for the perturbation (\ref{eq'}) is then explicitly solved by (see the appendix),  
\beq
\phi_1=- \frac{4}{9} L^4_* A^3 \sqrt{\frac{\pi}{3}} \(\frac{8}{a}\frac{\theta(r+t)}{(t-r)^3 r}+\frac{10a}{3}\frac{\p_t\delta(r+t)}{(t-r)^3 r}-2 a\frac{\delta(r+t)}{(t-r)^4 r}-\frac{16a}{5}\frac{\theta(r+t)}{(t-r)^5 r} \),
\eeq
which for  $ r >> a$,  reduces to 
\beq
\phi_1\(r \gg a \) =-\frac{32}{9} \sqrt{\frac{\pi}{3}}  L^4_* A^3 \frac{\theta(r+t)}{a(t-r)^3 r}.
\label{exact'}
\eeq
Notice,  that the latter expression includes an extra factor of $-4/3$ as compared to the 
solution (\ref{phione}) obtained by the averaging of the source.   This is not surprising, given the 
fact that the collapsing pulse moves at the speed of light, and the approximation of an instant  averaging should give an order-one correction\footnote{This can be understood as a consequence of  the  relations that hold (for  $t<0$) for any well-behaved function $Q(x)$ , 
 \begin{equation}
 4 {\delta(r+t) \over r}\left ( {dQ(x) \over dx} \right ) _{x = -2r} \, = \, \Box { \theta(r+t)  Q(t-r) \over r}  
  \label{aaa}
  \end{equation}
  and 
  \begin{equation}
  4 \pi Q(t)  \delta (\vec{r}) \, = \, \Box  {Q(t-r) \over r}  \, .
  \label{bbb}
  \end{equation}
  This  shows that an averaged source  $Q(t)\delta(\vec{r})$  that gives the same  $r> -t$ wave, 
differs from the exact one by a factor $ c \, = \,  4 t d_xQ|_{x=2t}  \, Q(t)^{-1}$.  In our case, 
$Q(x) \propto 1/x^3$  and thus, $c =   - 3/4$. }.

 
 The solution (\ref{exact'}) shows that  a collapsing  spherical wave leaves a wake of radial  "electric"  field $ \partial_j \phi$ and therefore classicalizes. 
 
 \section{Estimate of  $r_*$-radius}
 
 We shall now estimate the  $r_*$-radius of the classicalon.   
 The most straightforward estimate comes from the condition that  for $r\sim r_*$, the expansion parameter $\epsilon$, defined in (\ref{parameter}), becomes order one, 
   \begin{equation}
   \epsilon_* \, \equiv \, A^2L_*^4/(ar_*^3) \sim 1 \, , 
   \label{parameterstar}
\end{equation}   
which gives, 
   \begin{equation}
   r_* \, \sim  \,L_* (A^2L_*/a)^{1/3} \, . 
   \label{parameter'}
\end{equation}   
Translating this expression in terms of the center of mass energy 
 \begin{equation}
\sqrt{s} \, \sim \, {A^2 \over a} \,,
 \label{centere}
 \end{equation}
 we get 
 \begin{equation}
  r_*(s)  \, \sim \, L_*(L_* \sqrt{s})^{{1\over 3}} \, , 
 \label{starmax}
 \end{equation}
 which reproduces the estimate of \cite{dggk}.
 
  The  above estimate agrees with the following one:
 the radius  $r_*$ is defined as the maximal value of $r$ at which for  the  softest  (with maximal $a \sim r$) wave-packet the gradient of $\phi$ becomes of order $M_*^2$.  
  
   Indeed, consider a field $\phi_1$ produced by self-sourcing of the  wave-packet $\phi_0$ localized within the sphere 
 of radius $r $.  Outside the sphere the field is given by  (\ref{exact'}), its gradient being of order 
 \begin{equation}
  \partial_r \phi_1  \, \sim \,  {A^3 L_*^4\over ar^5}  \, .
 \label{electricradial}
 \end{equation}
 Equating this to $M_*^2$, we get,  
  \begin{equation}
r \, \sim \,  L_*\left ({A^3 L_* \over a} \right )^{{1\over 5}} \, .
 \label{star}
 \end{equation}
Translating this value in terms of the center of mass energy $\sqrt{s}$, given by 
(\ref{centere}), we get,   
 \begin{equation}
  r   \, \sim \,  L_* (\sqrt{s^3} L_*^2a)^{{1\over 10}} \, .
 \label{starstar}
 \end{equation}
 $r_*$ is given by the value of $r$ in the above expression, for which the spread $a$ is maximal, that is, $a \sim r$. 
 This gives  (\ref{starmax}).
 
  The equation (\ref{starmax})  confirms the outline of \cite{dggk} for the scattering process. 
  Since we are not solving the full scattering problem, we can only give qualitative estimates 
of the outcome, but relation (\ref{starmax}) gives us a very important information.   
  This relation indicates, that the size of classical configuration grows with center of mass energy, 
  implying that  few-to-few particle scatterings can only go through  very low momentum-transfer $\sim  M_*(L_* \sqrt{s})^{-{1\over 3}}$, and thus have an amplitude
$ \sim  (L_* \sqrt{s})^{-{4\over 3}} $.  For $\sqrt{s} \, \gg \, M_*$,  this is a negligible contribution into the scattering cross section, which has to be dominated by many-particle production through the decay of the classical configuration, and can be estimated to have a geometric cross-section, $\sigma \sim 
L_*^2(L_* \sqrt{s})^{{2\over 3}}$.

 \section{Importance of Energy-Sourcing} 
 
    As in the case of  monochromatic  waves,   the energy-sourcing is a defining property  
 for classicalization of localized wave-packets.    
     We wish to stress,  that  classicalization is {\it not}  equivalent to a standard dispersion of a wave-packet in a generic self-interacting theory.    For classicalization to take place, it is absolutely essential that the field $\phi$ is (self)sourced  by the energy. 
  For other types of self-interactions, the wave packet can still spread at short distances, but 
  because the $r_*$-radius  diminishes with energy, the 
  system  will not classicalize in general.  As an example, consider  the situation in which we replace the
 derivative  self-coupling  in (\ref{nambu}) by some other non-derivative self-interaction \footnote{We thank Alexander Pritzel and  Nico Wintergerst for motivating  clarification of this question 
 by their numerical studies.}
  \begin{equation}
 \mathcal{L}= {1\over 2}  \(\partial_{\mu}\phi\)^2 \, - \,  {L_*^{n-3} \over (n+1)}  \phi^{n+1} \, .
 \label{nambunew}
  \end{equation}
   The equation now becomes,  
    \begin{equation}
   \Box \phi \,  = \, -\,  L_*^{n-3}  \ \phi^{n} \, .
 \label{nambunewequ}
  \end{equation} 
  Applying the same perturbative expansion as before,  $\phi \, = \, \phi_0\, + \, \phi_1 \, +...$, and taking 
  $\phi_0$  in form of the collapsing  wave (\ref{wavepacket}), we get the following equation for 
  $\phi_1$, 
    \begin{equation}
   \Box \phi_1 \,  = \, -\,  A^n L_*^{n-3} {1 \over r^n}  e^{-n{(r+t)^2 \over a^2}} \, .
 \label{n_sourcing}
  \end{equation} 
 In the approximation of small $a$, this equation is solved by,  
   \begin{equation}
\phi_1 \,  = - \frac{(-2)^{n-1}}{4(n-2)} \sqrt{\frac{\pi}{n}} \,  ( A^n L^{n-3}_* a)\,  \frac{\theta(r+t)}{(t-r)^{n-2} r} \,.
 \label{phione111}
 \end{equation}
This solution indicates that the system does not classicalize, since  the $r_*$-radius is always 
below $L_*$.   Indeed,  the conditions 
$\phi_1 \sim \phi_0$  and  $\phi_1 \, \sim \, M_*$ are achieved for
 \begin{equation}
  r_*  \, \sim \,  L_* \,  A^{{n -1 \over n-2}} \left({a \over L_*}\right )^{{1\over n-2}} \, 
 \label{cond1}
 \end{equation}
and 
 \begin{equation}
  r_*  \, \sim \,  L_*  A^{{n \over n-1}} \left({a \over L_*}\right )^{{1\over n-1}} \, 
 \label{cond2}
 \end{equation}
 respectively.   Notice, that for $n=3$, the equation (\ref{cond1}) reproduces the result 
 (\ref{rzero}) of monochromatic wave-scattering in $\phi^4$-theory, if we replace $\omega \rightarrow  a^{-1}$ and 
 $L_*^{n-3}  \rightarrow \lambda$.  

   Now remembering that for a localized wave-packet  $a$ cannot exceed $r$, we 
 conclude that for $A \sim 1$, both conditions give $r_* \, \ll \, L_*$. Thus, the system does not classicalize.   
 
 
   The  above consideration  illustrates the importance of the energy-sourcing for classicalization, and thus the special role of derivatively-coupled scalars, such as Goldstone bosons.

\section{Classicallization of DGP-Galileon}

 Another  example we wish to consider  is  given by the action, 
  \begin{equation}
 \label{saction}
\mathcal{L} =(\partial_{\mu}\phi)^2 \left({1 \over 2} \, + \,  {L_*^3 \over 4}~ \Box \, \phi \right) \, + \,  \phi  \, J .
\end{equation}   
 This theory describes  the  self-interaction of  helicity-zero (St\"uckelberg) component of the 
DGP graviton, and can be viewed as the decoupling limit \cite{lpr} of that theory,  in which both the graviton
Compton wavelength as well as Planck scale are sent to infinity, whereas the scale
$L_*$ is kept fixed.  We shall treat the above theory in its own right, irrespective of 
its gravitational origin.   The remnant of gravitational nature, however, does persist 
in form of a residual symmetry under Galilean transformations, under which derivative of the scalar shifts by a constant  
\begin{equation}
\partial_{\mu} \phi \, \rightarrow \,\partial_{\mu} \phi +  c_{\mu}
\label{Galilei}
\end{equation}
 Because of this, such scalars are sometimes  referred to as {\it Galileons} \cite{nr}. 

 Because,  this theory is known to exhibit $r_*$-phenomenon \cite{lpr, dhk, nr, ddgv, gruzinov},   both for static as well as for the 
 time-dependent external sources,  in \cite{dggk} it was  identified as an interesting  candidate  
for classicalization.   After briefly reviewing the static argument, we shall extend the analysis 
to the scattering process.  We shall discover, that in scattering of a spherical wave, classicalization 
does not happen in  the leading order in non-linearities.  In the other words, the  flux of 
the $\phi$-gradient  can only form after the back-reaction on a linearized  wave pulse is taken into the account.  The  origin of this phenomenon is in the Galilean symmetry. 
 
   Before going to scattering case, let us briefly review classicalization of the theory by an external 
   source. 
    Consider a localized source at scale  $L$ and energy $\sqrt{s} \, = \, 1/L$.  At distances  $r \gg L$ such a source can be approximated by $4\pi \delta(\vec{r})\sqrt{s}L_*$.   The equation of motion 
   \begin{equation}
   \Box \, \phi + \, {L_*^3 \over 2} \left[ (\Box \, \phi)^2-(\partial_{\mu}\partial_{\nu} \phi)^2 \right] \, = \, J \, 
   \label{dgpeq}
   \end{equation}
   for a spherically symmetric ansatz
has an exact solution \cite{nr}, 
\begin{equation}
\partial_r\phi \, = \, {r \over 2L_*^3} \left( 1\pm \sqrt{1+\frac{4r_*^3}{r^3}}\right)
 \label{exact}
\end{equation}
where $r_* \equiv \,  L_* (L_*\sqrt{s})^{1/3}$.  Thus,   at distances  $r \, \gg \, r_* $
we have $\phi \, \sim \, L_*\sqrt{s} /r$,  whereas for $r \ll r_*$  we have $\phi \, \sim \, \sqrt{r\sqrt{s}}\, L_*^{-1}$. 
Thus, $r_*$ plays a role similar to the Schwarzschild radius and is the scale  at which $\phi$ classicalizes. ( An analogous effect in massive gravity was observed by Vainshtein in
\cite{arkady}.) 
   
 
We shall now abandon the external source and consider the situation of dynamical self-sourcing, calculating the correction to the zeroth order collapsing spherical wave-packet of the form (\ref{wavepacket}), following the above-mentioned procedure.  It is easy to see that the integrated source at this order is zero.  That is, 
\begin{equation}
\left [ \frac{L^3_*}{2} ~ \int d^3 x ~ \partial_t \left ( \frac{1}{2}\p_t(\p_\mu\phi_0)^2-\p_t\phi_0\Box\phi_0 \right ) \right ]_{\phi_0= \psi(r+t)/r}\, \, = \, 0 ,
\label{zerocharge }
\end{equation}
so that the self-sourcing can only appear after the correction to the linearized wave-solution from non-linearity is taken into account.   This result is confirmed by the explicit solution, which we obtain below. 
 
The equation of motion (\ref{dgpeq}) with $J=0$ yields that the first perturbation of a collapsing spherical wave-packet satisfies
\beq
\Box\phi_1&=&\frac{L^3_*}{2}\((\p_\mu \p_\nu \phi_0)^2 \) \nonumber \\ &=&\frac{L^3_*}{2}\(2(\p^2_t\phi_0)^2-2(\p_t\p_r\phi_0)^2-\frac{4}{r}\p^2_t\phi_0\p_r\phi_0 +6\(\frac{\p_r\phi_0}{r}\)^2\)\equiv \bar j(r,t),
\label{eq'1}
\eeq
where the zeroth order equation $$\Box\phi_0=(\p^2_t-\p^2_r-\frac{2}{r}\p_r)\phi_0=0$$ has been used. 
Using the explicit form of the zeroth order wave-packet (\ref{zerophi}), the source $\bar j$ rewrites as
\beq
\bar j(r,t)=L^3_*\(\frac{2\psi ''\psi}{r^4}+ \frac{2\psi '^2}{r^4}-\frac{6 \psi '\psi}{r^5}+\frac{3\psi^2}{r^6}\).\nonumber
\eeq
Evaluating the last expression for $\psi$ given by (\ref{wavepacket}), and then taking the $a\to 0$ limit, it can be recast in the following form
\beq
\bar j(r,t)=L^3_* A^2\sqrt{\frac{\pi}{2}}\(\frac{a}{r^4}\p^2_r-\frac{3 a}{r^5}\p_r+\frac{3 a}{r^6} \)\delta(r+t).
\eeq
Unlike the previous case, the leading $1/a$ contribution cancels for the DGP-Galileon. Moreover, solving (\ref{eq'1}) by the methods given in the appendix, we obtain
\beq
\phi_1=-L^3_* A^2\sqrt{\frac{\pi}{2}} a \frac{\p_t\delta(r+t)}{(t-r)^2 r}.
\eeq
Thus the spread of a DGP-Galileon wave-packet does not happen at this  order in non-linearity. 
 This difference as compared to the Goldstone case can be traced to the Galilean symmetry 
 (\ref{Galilei})  of the Lagrangian (\ref{saction}). 

\section{Conclusions}

  In this note we have studied dynamics of classicalization phenomenon suggested in \cite{dggk}.  
   For understanding the fundamental difference between the classicalizing theories and the  ones 
   that do not exhibit such an effect, we have  generalized the notion of classical
   $r_*$-radius to the latter class of theories.   An universal physical meaning of 
   $r_*$-radius  can be defined as of a shortest distance down to which, in a scattering process,  particles propagate freely, without experiencing a significant  
  interaction.  By default, the scattering cross section then emerges as a geometric cross section set by $r_*$, given by (\ref{unisection}). The difference between classicalizing and non-classicalizing theories then can be traced to different  behaviors 
 of  $r_*$-radius with the growing energy. 
   The defining property of  classicalizing theories is the growth of  $r_*$-radius 
 with energy.  
  
    As an example,  we investigated the system of a derivatively self-coupled scalar, that serves as a prototype 
   for  Nambu-Goldstone-St\"uckelberg  field.  In order to study classicalization in a simple
   scattering process, we have prepared an initial collapsing $S$-wave-packet, with high energy 
   but small occupation number.  In this way,  at the initial stage system can be regarded as being in a quantum state with small number of particles with very high center of mass energy. 
    We then observed that as  the wave collapses the classical configuration gradually develops   because of the self-sourcing of the Goldstone field by its own energy.   Self-sourcing (or sourcing) by energy is a crucial  factor for classicalization, which makes it different from a simple dispersion of the wave-packet in ordinary interacting theories.  Increase of energy  inevitably produces a growing self-source  of the classicalon  field, and leads to the creation of a classical configuration of $r_*$-radius which confirms the estimate of \cite{dggk} for the analogous scattering process.   
    
     Classicalon is {\it not}  created as a result of quantum transition, 
but is developed gradually because of sourcing by the same energy that makes particles scatter.  
 Because of this fact,  production of classicalons, in sharp difference from non-perturbative solitons, is not only unsuppressed, but in fact dominates the scattering process at 
trans-cutoff  energies.  
  
     We have performed analogous study for the theory of DGP-Galileon, and discovered that in 
     the leading order in self-sourcing the classicalization does not happen.  This suppression can be  traced to a high-symmetry of the model.  
     
     Our analysis indicates that the nature of  (self)sourcing  crucially determines the outcome 
     of classicalization, and that Goldstone fields appear as interesting candidates  for this 
     phenomenon.

  \vspace{5mm}
\centerline{\bf Acknowledgments}
We thank Gian Giudice, Cesar Gomez and Alex Kehagias  for  valuable  on-going discussions on classicalization.  We thank Lasha Berezhiani for discussions on  classicalization in other systems. 
 The work of G.D. was supported in part by Humboldt Foundation under Alexander von Humboldt Professorship,  by European Commission  under 
the ERC advanced grant 226371,  by  David and Lucile  Packard Foundation Fellowship for  Science and Engineering, and  by the NSF grant PHY-0758032. D.P. is supported by the Mark Leslie Graduate Assistantship at NYU.

\section*{Appendix A. Explicit Solution}
In this appendix we give a detailed derivation of classicalization for a localized wave-packet in the self-interacting theory, defined by the lagrangian (\ref{nambu}).

We start with the equation of motion for the $\phi$-field
\beq
\Box \phi =-L^4_*~ \p^\mu (\p_\mu \phi (\p_\nu \phi)^2),
\eeq
and build up the perturbation series in the small parameter $L^4_*$
\beq
\phi=\phi_0 + \phi_1 + ...~.
\eeq
The first perturbation $\phi_1$ on the solution of the free theory is sourced by the localized source $j(r,t)$, composed of the derivatives of the initial-state wave-packet
\beq
\Box \phi_1 = -L^4_*~ \p^\mu (\p_\mu \phi_0 (\p_\nu \phi_0)^2)= j(r,t).
\label{eq}
\eeq
Using the explicit form of the zeroth-order solution (\ref{zerophi}), the source can be rewritten in the following way
\beq
 j(r,t)=-L^4_* \left( 2 \frac{\psi^2\psi ''}{r^5}+8 \frac{\psi\psi '^2}{r^5}-12 \frac{\psi^2\psi '}{r^6}+4 \frac{\psi^3}{r^7} \right).\
\label{source}
\eeq
Let us take the function $\psi$ in the form of a gaussian $\psi=A e^{-\frac{(r+t)^2}{a^2}}$, so that the last expression becomes

\beq
j(r,t)=- L^4_* A^3 \( \frac{8}{3 a^2 r^5} -\frac{4}{r^6}\p_r +\frac{10}{9r^5} \p^2_r +\frac{4}{r^7}\) e^{-\frac{3(r+t)^2}{a^2}}.
\eeq
Taking the $a\to 0$ limit and using the relation 
\beq
\lim_{a\to 0} e^{-\frac{3(r+t)^2}{a^2}}=a\sqrt{\frac{\pi}{3}}\delta(r+t),
\eeq
the source rewrites as
\beq
j(r,t)=- L^4_* A^3 \sqrt{\frac{\pi}{3}} \(\frac{8}{3 a r^5} -\frac{4a}{r^6}\p_r +\frac{10a}{9r^5} \p^2_r  +\frac{4a}{r^7} \) \delta(r+t).
\label{source1}
\eeq
We shall solve the laplace equation for each of the four parts of the right hand side of the last expression viewed as separate sources. The full solution will be the superposition of the four waves
\beq
\phi_1=\phi_1^{(1)}+\phi_1^{(2)}+\phi_1^{(3)}+\phi_1^{(4)}.
\eeq
We start with the equation for $\phi_1^{(1)}$, sourced by the first term in (\ref{source1})
\beq
\Box\phi_1^{(1)}=- L^4_* A^3 \sqrt{\frac{\pi}{3}}\frac{8}{3 a} \frac{\delta(r+t)}{r^5}.
\eeq
The last equation can be directly solved via the relation
\beq
\Box^{-1}\( \frac{\delta(r+t)}{r^{n+2}} \)=\frac{(-2)^{n+1}}{4n}\frac{\theta(r+t)}{(t-r)^n r},
\eeq
which leads to the following expression for $\phi_1^{(1)}$,
\beq
\phi_1^{(1)}=-L^4_* A^3 \sqrt{\frac{\pi}{3}} \frac{32}{9a}\frac{\theta(r+t)}{(t-r)^3 r}.
\eeq
The latter expression represents the dominant contribution to the spread of the wave-packet at distances $r\gg a$.

Analogously, the field sourced by the second term in (\ref{source1}), satisfies the following equation
\beq
\Box\phi_1^{(2)}=L^4_* A^3 \sqrt{\frac{\pi}{3}} 4 a  \frac{\p_r \delta(r+t)}{r^6}=L^4_* A^3 \sqrt{\frac{\pi}{3}} 4 a  \frac{\p_t \delta(r+t)}{r^6}.
\eeq
Defining a new function $f_1^{(2)}$ through $\phi_1^{(2)}\equiv\p_t f_1^{(2)}$, we have
\beq
\Box f_1^{(2)}=L^4_* A^3 \sqrt{\frac{\pi}{3}} 4 a  \frac{\delta(r+t)}{r^6}+g(r),
\eeq
where $g(r)$ is an integration 'constant', which produces a static configuration for $f^{(2)}_1$ and is therefore irrelevant for the expression for $\phi^{(2)}_1$, so that we can discard it in what follows. The last equation can now be solved 
\beq
f_1^{(2)}=-L^4_* A^3 \sqrt{\frac{\pi}{3}} 8a \frac{\theta(r+t)}{(t-r)^4r}.
\eeq
Differentiating by time, we obtain the expression for $\phi^{(2)}_1$
\beq
\phi^{(2)}_1=-L^4_* A^3 \sqrt{\frac{\pi}{3}} \(8a \frac{\delta(r+t)}{(t-r)^4r}-32a \frac{\theta(r+t)}{(t-r)^5 r}\).
\eeq

For finding $\phi^{(3)}_1$, we have to solve the following equation
\beq
\Box \phi^{(3)}_1= -L^4_* A^3 \sqrt{\frac{\pi}{3}} \frac{10 a}{9}\frac{\p_r^2\delta(r+t)}{r^5}=-L^4_* A^3 \sqrt{\frac{\pi}{3}} \frac{10 a}{9}\frac{\p_t^2\delta(r+t)}{r^5}.
\eeq
Let us again introduce an auxiliary field $\phi_1^{(3)}\equiv\p^2_t f_1^{(3)}$, which satisfies
\beq
\Box f^{(3)}_1= -L^4_* A^3 \sqrt{\frac{\pi}{3}} \frac{10 a}{9}\frac{\delta(r+t)}{r^5}
\eeq
and is therefore given by
\beq
f^{(3)}_1= -L^4_* A^3 \sqrt{\frac{\pi}{3}} \frac{40 a}{27}\frac{\theta(r+t)}{(t-r)^3 r}.
\eeq
Taking the second time derivative, we obtain
\beq
\phi^{(3)}_1=-L^4_* A^3 \sqrt{\frac{\pi}{3}}\(\frac{40a}{27}\frac{\p_t \delta(r+t)}{(t-r)^3 r}-\frac{80a}{9}\frac{\delta(r+t)}{(t-r)^4 r}+\frac{160a}{9}\frac{\theta(r+t)}{(t-r)^5 r}\).
\eeq
The fourth contribution $\phi^{(4)}_1$ can be directly found from its equation of motion
\beq
\phi^{(4)}_1=-L^4_* A^3 \sqrt{\frac{\pi}{3}}\frac{64a}{5}\frac{\theta(r+t)}{(t-r)^5r}.
\eeq
Superimposing all the contributions, we obtain the exact expression for the first perturbation on the localized wave-packet
\beq
\phi_1=- \frac{4}{9} L^4_* A^3 \sqrt{\frac{\pi}{3}} \(\frac{8}{a}\frac{\theta(r+t)}{(t-r)^3 r}+\frac{10a}{3}\frac{\p_t\delta(r+t)}{(t-r)^3 r}-2 a\frac{\delta(r+t)}{(t-r)^4 r}-\frac{16a}{5}\frac{\theta(r+t)}{(t-r)^5 r} \).
\eeq

\newpage

\end{document}